\documentclass[prb,twocolumn,showpacs,bm,amssymb,amsmath,floatfix]{revtex4}
\usepackage{epsfig}
\begin{document}
\title{Charge-density-wave order parameter of the Falicov-Kimball model
in infinite dimensions}
\date{\today}
\author{Ling Chen}
\email{lc63@georgetown.edu}
\affiliation{Department of Physics, Georgetown University, Washington, DC
20057}
\author{B. A. Jones}
\email{bajones@almaden.ibm.com}
\affiliation{IBM, Almaden Research Center, 650 Harry Rd., San Jose, CA, 95120}
\author{J. K. Freericks}
\email{freericks@physics.georgetown.edu}
\homepage{http://www.physics.georgetown.edu/~jkf}
\affiliation{Department of Physics, Georgetown University, Washington, DC
20057}

\begin{abstract}
In the large-$U$ limit, the Falicov-Kimball model maps onto an effective
Ising model,
with an order parameter described by a BCS-like mean-field theory in infinite
dimensions.  In the small-$U$ limit, van Dongen and Vollhardt showed that the
order parameter assumes a strange non-BCS-like shape with a sharp reduction
near $T\approx T_c/2$.  Here we numerically investigate the crossover between
these two regimes and qualitatively determine the order parameter for a 
variety of different values of $U$. We find the overall behavior of the 
order parameter as a function of temperature to be quite anomalous.
\end{abstract}

\maketitle

Dynamical mean field theory\cite{metzner_vollhardt_1989}
has been widely used to study electron correlations
in a variety of different interacting systems.  Much work has focused on the
``paramagnetic'' metal-insulator transition, on transport properties in
the normal state, and on determining phase diagrams to ordered phases via
a susceptibility analysis or a Maxwell construction.  The properties of the
ordered phase have been less studied, yet there is much interesting physics
to examine there.  For example, one might have thought that since the
system is infinite-dimensional, both the critical behavior and the order
parameter as a function of temperature would be determined by a BCS-like
mean-field picture.  Indeed, the critical exponents are always
mean-field like, but the
order parameter can have quite anomalous behavior as a function of temperature.
This anomalous
behavior is amplified for small correlation strength, since many models
map onto effective spin models for large correlations, and the order parameter
of an infinite-dimensional
spin model is always mean-field-like.  In this contribution, we examine in
detail the case of a commensurate (two-sublattice) charge-density-wave (CDW)
phase in the spinless Falicov-Kimball model at half filling.  Analytical work
in the large and small-$U$ limits has already been carried 
out\cite{vandongen_vollhardt_1990,vandongen_1992}.

The Falicov Kimball (FK) model\cite{falicov_kimball_1969} was
introduced in 1969 to describe metal-insulator transitions in
transition-metal and rare-earth compounds. The model consists of two types of
particles: itinerant conduction electrons and localized ions that mutually 
interact with an on-site Coulomb interaction.  It
is the simplest Fermionic model of electron correlations and the spinless
version can be interpreted as a simple model for 
crystallization,\cite{kennedy_lieb_1986} where
the system has a phase transition from a disordered phase
at high temperature to an ordered phase as the temperature
is lowered.  We examine the order parameter as a function of temperature here.

Many body effects
enter via the statistical mechanics associated with annealed
averaging. The FK model is the simplest many-body problem that can be solved
exactly in the limit of large dimensions. Brandt and
Mielsch\cite{brandt_mielsch_1989,brandt_mielsch_1990,brandt_mielsch_1991} 
presented the first solution of this problem using
dynamical mean-field theory. Their solution quantitatively illustrated how a
period-two CDW phase is stabilized at low
temperatures.

For the symmetric half-filled case on a bipartite lattice, 
the ordering is into a
commensurate CDW state, in which the particles order in a
checkerboard pattern: the conduction electrons preferentially occupy one
sublattice, and the ions the other one. Van Dongen and 
Vollhardt\cite{vandongen_vollhardt_1990,vandongen_1992} derived an analytical 
expression for the critical temperature $T_{c}$ 
as a function of the Coulomb interaction in the large and small-$U$ limits.
For large-$U$, it has the conventional strong-coupling $t^2/U$ form, but
for small coupling, $T_c$ is much larger than the BCS-like exponential
form of $\exp[-C/U]$.
They obtained similar results for the order parameter---for large-$U$ it
behaved mean-field-like, but for small-$U$ the behavior was quite anomalous
with a sharp decrease as $T\rightarrow T_c/2$.
Here we numerically explore the behavior of the order
parameter between these two limits. A self-consistent 
algorithm\cite{gruber_macris_royer_2001}
is employed to perform efficient computations at low temperature and to
make it feasible to study the order parameter when $U$ is small.
Surprisingly, the order parameter is not BCS-like!

The spinless Falicov-Kimball model is represented by the following
Hamiltonian:
\begin{equation}
\mathcal{H}=-\sum_{ij}t_{ij}c^\dagger_ic_j+E_f\sum_if^\dagger_if_i+U
\sum_ic^\dagger_ic_if^\dagger_if_i.
\label{eq: ham}
\end{equation}
The conduction electrons (created or destroyed at site \textit{i} by
$c^\dagger_i$ or $c_i$) can hop between nearest-neighbor sites
with a hopping matrix\cite{metzner_vollhardt_1989}
$-t_{ij}=-t^*/2\sqrt{D}$. The localized ions (created or
destroyed at site \textit{i} by $f^\dagger_i$ or $f_i$) have a site energy
$E_f$. There is a Coulomb interaction $U$ between the localized ions
and the conduction electrons that sit at the same lattice site. A chemical
potential $\mu$ is employed to determine the number of conduction
electrons.  At half filling, we have $\mu=U/2$ and $E_f=-U/2$.

We examine two bipartite lattices here---the hypercubic lattice with a 
density of states (DOS) given by 
$\rho(\epsilon)=\exp (-\epsilon^2)/t^*\sqrt{\pi}$ and the infinite-coordination
Bethe lattice with $\rho(\epsilon)=\sqrt{4-\epsilon^2}/t^*2\pi$ (we take
$t^*=1$ as our energy unit).
The local Green's function $G(i\omega_n)$ can be
written as the Hilbert transform of the noninteracting DOS
$\rho(\epsilon)$
\begin{equation}
G(i\omega_n)=\int
d\epsilon\rho(\epsilon)\frac{1}{i\omega_n+\mu-\Sigma_n-\epsilon}
\label{eq: g_hilbert}
\end{equation}
where $\omega_n=\pi T(2n+1)$ is the Fermionic Matsubara frequency
and $\Sigma(i\omega_n)=\Sigma_n$ is the local self energy. 
Dyson's equation for the local self energy reads
\begin{equation}
\Sigma_n=i\omega_n+\mu-\lambda_n-G^{-1}(i\omega_n) 
\label{eq: self}
\end{equation}
with $\lambda_n$ the dynamical mean field evaluated at the $n$th Matsubara
frequency. The effective medium
Green's function $G_0(i\omega_n)$ is
\begin{equation}
G_0(i\omega_n)=\frac{1}{G^{-1}(i\omega_n)+\Sigma_n}=
\frac{1}{i\omega_n+\mu-\lambda_n}.
\label{eq: g0def}
\end{equation}
Solving the atomic problem in a time-dependent field yields
another equation for the local Green's function
\begin{equation}
G(i\omega_n)=\frac{1-w_1}{i\omega_n+\mu-\lambda_n}+
\frac{w_1}{i\omega_n+\mu-\lambda_n-U}
\label{eq: g_atomic}
\end{equation}
where $w_1$ represents the density of the ions
which is determined by the atomic partition function
\begin{equation}
\mathcal{Z}_{at}(\lambda)=\mathcal{Z}_0(\lambda,\mu)+e^{-\beta
E_f}\mathcal{Z}_0(\lambda,\mu-U)
\end{equation}
with
\begin{equation}
\mathcal{Z}_0(\lambda,\mu)=2e^{\frac{\beta\mu}{2}}\prod_n\frac{i\omega_n+\mu-\lambda_n}{i\omega_n}
\end{equation}
and $w_1={e^{-\beta E_f}\mathcal{Z}_0(\lambda,\mu-U)}/{\mathcal{Z}_{at}}$.
When $E_f=-U/2$  and $\mu=U/2$, then  $\lambda_n$ is pure imaginary
and we have $w_1=0.5$.

The iterative algorithm for finding the 
self-energy is that of Jarrell\cite{jarrell_1992}: (i) first
set the self-energy equal to zero; (ii) then determine the local Green's
function from Eq.~(\ref{eq: g_hilbert}); (iii) then
determine the effective-medium Green's
function from Eq.~(\ref{eq: g0def}) and the ion density from
the atomic partition function (in our case we always have $w_1=0.5$); (iv) next
determine the new local Green's function from Eq.~(\ref{eq: g_atomic}) and
the new self energy from Eq.~(\ref{eq: self}).
Starting with the new self-energy, repeat steps (ii-iv) until convergence
is reached.

\begin{figure}[htb]
\epsfxsize=3.0in
\epsffile{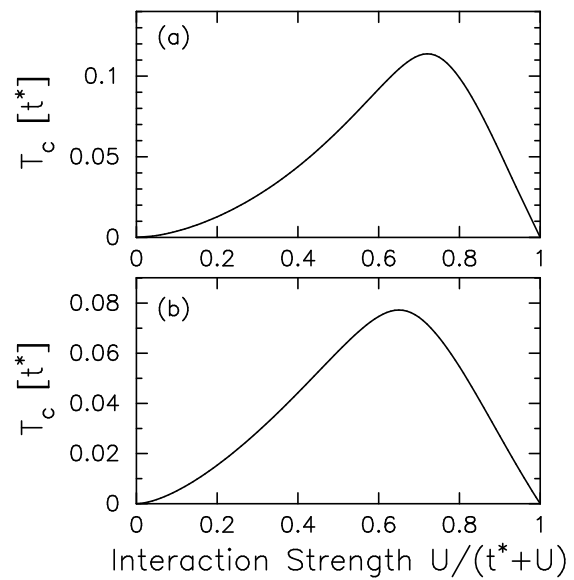}
\caption{\label{fig: tc} Transition temperature for the $(X=-1)$ CDW at
half filling on (a) the Bethe lattice
and (b) the hypercubic lattice as determined from the solution of 
Eq.~(\ref{eq: tc_eq}).  The horizontal axis is plotted as $U/(t^*+U)$ in order
to include both the weak and strong coupling results on the same graph.}
\end{figure}

After solving for the self-energy and the Green's functions, the
susceptibility to a CDW instability
($T\chi=\langle n_in_j\rangle-\langle n_i\rangle\langle n_j\rangle$) can be 
calculated following the
derivation of Brandt and Mielsch\cite{brandt_mielsch_1989,brandt_mielsch_1990}.
At high temperature, the electrons are uniformly distributed
throughout the lattice. As the temperature is lowered, the charge
density becomes nonuniform when the momentum-dependent
susceptibility diverges and a CDW forms. 
After some tedious algebra, the transition temperature is found to occur when
\begin{eqnarray}
1&=&\sum_n\frac{w_1(1-w_1)U^2G^3_n\eta_n(q)}
{(1+G_n\Sigma_n)[1+G_n(\Sigma_n-U)]}\cr
&\times&\frac{1}{1+G_n[\Sigma_n-(1-w_1)U]+
G_n\eta_n(q)[1+G_n(2\Sigma_n-U)]}
\label{eq: tc_eq}
\end{eqnarray}
holds, where
\begin{equation}
\eta_n(q)=-\frac{G_n}{\tilde{\chi}^0_n(q)}-G_n^{-1}
\end{equation}
and the bare susceptibility satisfies
\begin{equation}
\tilde{\chi}^0_n(q)=-\frac{1}{N}\sum_qG_n(k+q)G_n(q)
\end{equation}
in terms of the momentum-dependent Green's functions ($N$ is the number of
lattice sites).
The q-dependence of $\chi$ can be summarized by the scalar 
parameter\cite{mueller-hartmann_1989b}
\begin{equation}
X(q)=\lim_{D\rightarrow\infty}\frac{1}{D}\sum_{i=1}^D \cos q_i.
\end{equation}
For $X=-1$, there is a phase transition from a uniform
distribution to a (two-sublattice) checkerboard phase when the temperature 
falls below $T_c$.  This transition temperature is plotted versus $U$ for
(a) the Bethe lattice and (b) the hypercubic lattice in Fig.~\ref{fig: tc}.

For $T<T_c$, we need to revise the algorithm to solve
for the Green's functions and to determine the order parameter. Since the 
thermodynamic quantities now differ on the $A$ and $B$ sublattices of the
bipartite lattice, we label them by their respective sublattice. The order
parameter for the CDW is defined to be $\Delta=|w_1^A-w_1^B|$.  We find that
if we try to iterate a generalization of the Jarrell algorithm to include 
the possibility of CDW order, then the equations suffer from ``critical
slowing down'' as one approaches $T_c$ from below.  This would make an
efficient determination of the order parameter as a function of $T$ impossible
when $T_c$ became small because the number of Matsubara frequencies
employed becomes large, and the equations must be iterated for a long
time to achieve convergence.  Instead, we adopt a variant of the algorithm
proposed by Gruber \textit{et al.}\cite{gruber_macris_royer_2001}.  We
start by setting the self energies on each sublattice to zero.  Then we
(i) choose a value for the order parameter $0<\Delta<1$ and set 
$w_1^A=0.5(1+\Delta)$ and $w_1^B=0.5(1-\Delta)$; (ii) we determine the parameter
$Z_n$ defined by
\begin{equation}
Z_n=\sqrt{(i\omega_n+\mu-\Sigma^A_n)(i\omega_n+\mu-\Sigma^B_n)}
\end{equation}
and calculate the local Green's function on the $A$ sublattice
\begin{equation}
G^A_n=\frac{i\omega_n+\mu-\Sigma^B_n}{Z_n}\int
d\epsilon\frac{\rho(\epsilon)}{Z_n-\epsilon};
\label{eq: hilberta}
\end{equation}
(iii) we solve for the effective medium on the $A$ sublattice
\begin{equation}
G^A_0(i\omega_n)=({G^A_n}^{-1}+\Sigma^A_n)^{-1};
\end{equation}
and (iv) we calculate the new local Green's function
\begin{equation}
G^A_n=(1-w^A_1)G^A_0(i\omega_n)+\frac{w^A_1}{{G^A_0}^{-1}(i\omega_n)-U};
\end{equation}
finally (v) we determine the new self energy
\begin{equation}
\Sigma^A_n={G^A_0}^{-1}(i\omega_n)-{G^A_n}^{-1}.
\end{equation}
Next we perform a similar analysis on the $B$ sublattice: (vi) first we
update $Z_n$ then calculate the local Green's function on the $B$-sublattice
\begin{equation}
G^B_n=\frac{i\omega_n+\mu-\Sigma^A_n}{Z_n}\int
d\epsilon\frac{\rho(\epsilon)}{Z_n-\epsilon};
\label{eq: hilbertb}
\end{equation}
(vii) we determine the effective medium
\begin{equation}
G^B_0(i\omega_n)=({G^B_n}^{-1}+\Sigma^B_n)^{-1};
\end{equation}
and (viii) we find the new local Green's function
\begin{equation}
G^B_n=(1-w^B_1)G^B_0(i\omega_n)+\frac{w^B_1}{{G^B_0}^{-1}(i\omega_n)-U};
\end{equation}
finally (ix) we calculate the new self energy
\begin{equation}
\Sigma^B_n={G^B_0}^{-1}(i\omega_n)-{G^B_n}^{-1}.
\end{equation}
Steps  (ii--ix) are repeated until converged.  Then we (x) calculate
the local ion site energy on each sublattice by solving the generalization of
the formula for the filling $w_1$ to the two-sublattice case, and solving
for the $E_f$ on each sublattice:
\begin{equation}
E^A_f=\frac{U}{2}+T\ln \left [ 
\frac{1-w^A_1}{w^A_1}\right ] + T \sum_n\ln [1-UG^A_0(i\omega_n)],
\end{equation}
and
\begin{equation}
E^B_f=\frac{U}{2}+T\ln \left [
\frac{1-w^B_1}{w^B_1}\right ] + T \sum_n\ln [1-UG^B_0(i\omega_n)].
\end{equation}
If $E_f^A=E_f^B$, then a consistent thermodynamic solution has been achieved.
If not, then we adjust the order parameter $\Delta$
and repeat the algorithm starting
at step (i) until a consistent solution is reached.  Since the ion densities
on each sublattice are not updated during the iterative part of the algorithm,
this technique does not suffer from critical slowing down, and it is easy
to perform on both the Bethe lattice and the hypercubic lattice, simply by
changing the corresponding DOS in the Hilbert transforms of 
Eqs.~(\ref{eq: hilberta}) and (\ref{eq: hilbertb}). In our numerical 
calculations, we use an energy cutoff of $10t^*$ for the Matsubara
frequencies; the maximum number of Matsubara frequencies we include
is 250~000.

For $U\rightarrow\infty$, the FK model maps onto an antiferromagnetic 
Ising model  with $J=t^{*2}/4DU$. This spin system can be
analyzed with static mean field theory to yield
the order parameter in a Curie-Weiss form,
\begin{equation}
\Delta=\tanh (\frac{\Delta T_c}{T}),
\end{equation}
with $T_c=t^{*2}/4U$ on the hypercubic lattice.
For $U\rightarrow 0$, the order parameter varies sharply from this
result, with a steep decrease near $T_c/2$ and a flattening out before
going to zero at $T_c$.  We examine the transition between these two
limits numerically.  We find that $U$ must be very small to even see
the deviations from the mean-field theory curve.  The evolution from
the small-$U$ to large-$U$ limits is nonmonotonic.  As $U$ increases, the
order parameter curve becomes steeper and steeper near $T_c$ and flatter and 
flatter for small $T$ until we reach the $U$ corresponding to approximately
the maximum of the $T_c$ versus $U$ curve.  Once $U$ increases 
past the maximum, then the order parameter curve slowly approaches 
the large-$U$ limit.

\begin{figure}[htbf]
\epsfxsize=3.0in
\epsffile{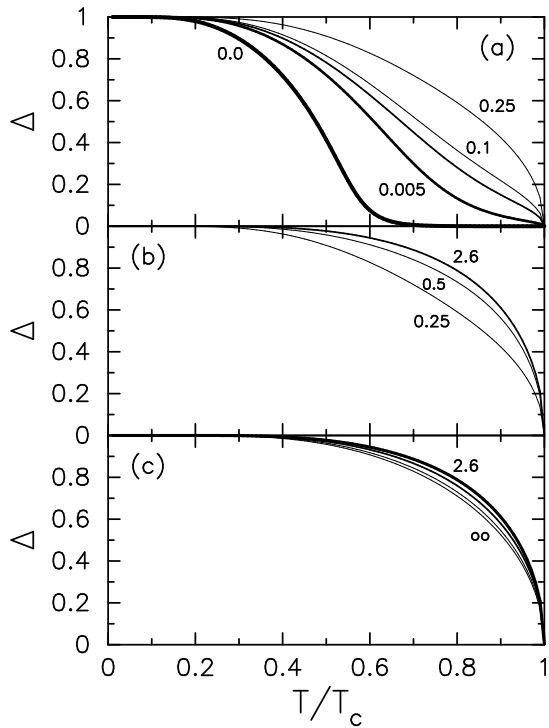}
\caption{\label{fig: bethe} Order parameter for the $(X=-1)$ CDW of the
FK model at half filling on the Bethe lattice.  Panel (a) shows the small-$U$
results where a depression can be seen ($U\rightarrow 0$, 0.005, 0.03, 0.1,
and 0.25), panel (b) shows the intermediate results ($U$=0.25, 0.5, and 2.6),
where the order parameter curve moves out to the right as $U$ is increased,
and panel (c) shows the large-$U$ regime ($U$=2.6, 3.0, 4.0 and $\infty$), 
where the curve shifts to the
left and ultimately takes the mean-field-theory form.}
\end{figure}

We illustrate this behavior in Fig.~\ref{fig: bethe} for the Bethe lattice.
When $U$ lies below 0.2 [panel (a)], we can see the depression in the curve 
develop as we evolve toward the $U=0$ limit.  For $0.2<U<2.6$, the curve moves 
the right and becomes steeper (but the critical exponent always remains 0.5)
[panel (b)]. Finally, for larger values of $U$, the curve moves to the left
until it assumes the mean-field-theory form [panel (c)].

\begin{figure}[htbf]
\epsfxsize=3.0in
\epsffile{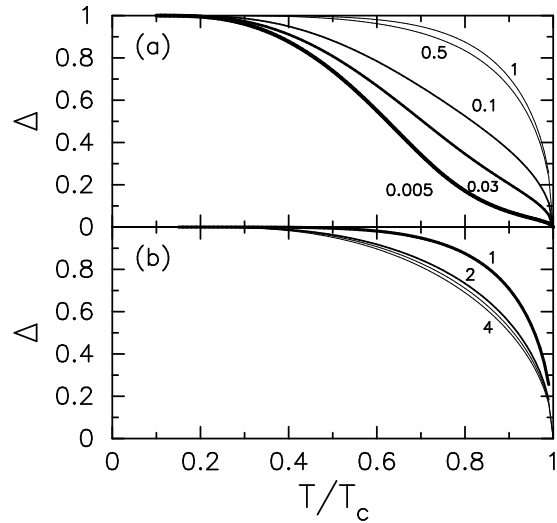}
\caption{\label{fig: hyper} Order parameter for the $(X=-1)$ CDW of the
FK model at half filling on the hypercubic lattice.  
Panel (a) shows the small-$U$
results where a depression can be seen ($U$= 0.005, 0.03, 0.1, 0.5, and 1.0) 
and panel (b) shows the larger-$U$ results ($U$=1.0, 2.0, 4.0, and $\infty$),
where the order parameter curve moves back to the left as $U$ is increased,
but overshoots, and then moves to the right to
ultimately take the mean-field-theory form.}
\end{figure}

The results on the hypercubic lattice are similar and are shown in
Fig.~\ref{fig: hyper}.  We show only two different regimes, as
the curve is steepest when $U$ lies near 1.0.
Note that there are no analytic results
available for the $U\rightarrow 0$ limit on the hypercubic lattice.
The one difference from the Bethe lattice, is that here, the curves move to
the right as we increase from $U=1$ to $U=4$, but they overshoot the
mean-field theory curve, so they move back to the left to reach the
$U\rightarrow\infty$ limit for larger values of $U$.

In summary, we have shown that the order parameter has an anomalous evolution 
as a function of temperature for small $U$ in the spinless FK model.  One
can ask is this anomalous behavior something generic to many-body systems,
to the infinite-dimensional limit, or to the FK model.  We tend to believe
that it is most likely a curious property of the FK model itself.  The
weak-coupling limit of the FK model is known to have anomalous behavior for
it's $T_c$, being much higher than what would be predicted by a BCS 
approach, and it is possible that the anomalous shape survives in finite
dimensions as well.  Unfortunately, our numerical calculations show that
$U$ must be very small in order to see the anomalous shape emerge, and it
is likely that the $T_c$ is so small, that numerical calculations in finite
dimensions may have trouble being performed at the temperatures necessary
to see the anomaly.
Even in infinite-dimensions, we are limited by how small we can 
make $U$ because the $T_c$
is pushed so low that the number of Matsubara frequencies included in
the calculation grows too large.

\textit{Acknowledgments}: We acknowledge support from the National
Science Foundation under grant number DMR-0210717.  L.C. would like
to thank the IBM Almaden Research Center for hosting him as a visitor 
while this work was completed.

\bibliography{fk_dmft.bib}

\begin{thebibliography}{11}
\expandafter\ifx\csname natexlab\endcsname\relax\def\natexlab#1{#1}\fi
\expandafter\ifx\csname bibnamefont\endcsname\relax
  \def\bibnamefont#1{#1}\fi
\expandafter\ifx\csname bibfnamefont\endcsname\relax
  \def\bibfnamefont#1{#1}\fi
\expandafter\ifx\csname citenamefont\endcsname\relax
  \def\citenamefont#1{#1}\fi
\expandafter\ifx\csname url\endcsname\relax
  \def\url#1{\texttt{#1}}\fi
\expandafter\ifx\csname urlprefix\endcsname\relax\def\urlprefix{URL }\fi
\providecommand{\bibinfo}[2]{#2}
\providecommand{\eprint}[2][]{\url{#2}}

\bibitem[{\citenamefont{Metzner and Vollhardt}(1989)}]{metzner_vollhardt_1989}
\bibinfo{author}{\bibfnamefont{W.}~\bibnamefont{Metzner}} \bibnamefont{and}
  \bibinfo{author}{\bibfnamefont{D.}~\bibnamefont{Vollhardt}},
  \bibinfo{journal}{Phys. Rev. Lett.} \textbf{\bibinfo{volume}{62}},
  \bibinfo{pages}{324} (\bibinfo{year}{1989}).

\bibitem[{\citenamefont{van Dongen and
  Vollhardt}(1990)}]{vandongen_vollhardt_1990}
\bibinfo{author}{\bibfnamefont{P.~G.~J.} \bibnamefont{van Dongen}}
  \bibnamefont{and}
  \bibinfo{author}{\bibfnamefont{D.}~\bibnamefont{Vollhardt}},
  \bibinfo{journal}{Phys. Rev. Lett.} \textbf{\bibinfo{volume}{65}},
  \bibinfo{pages}{1663} (\bibinfo{year}{1990}).

\bibitem[{\citenamefont{van Dongen}(1992)}]{vandongen_1992}
\bibinfo{author}{\bibfnamefont{P.~G.~J.} \bibnamefont{van Dongen}},
  \bibinfo{journal}{Phys. Rev. B} \textbf{\bibinfo{volume}{45}},
  \bibinfo{pages}{2267} (\bibinfo{year}{1992}).

\bibitem[{\citenamefont{Falicov and Kimball}(1969)}]{falicov_kimball_1969}
\bibinfo{author}{\bibfnamefont{L.~M.} \bibnamefont{Falicov}} \bibnamefont{and}
  \bibinfo{author}{\bibfnamefont{J.~C.} \bibnamefont{Kimball}},
  \bibinfo{journal}{Phys. Rev. Lett.} \textbf{\bibinfo{volume}{22}},
  \bibinfo{pages}{997} (\bibinfo{year}{1969}).

\bibitem[{\citenamefont{Kennedy and Lieb}(1986)}]{kennedy_lieb_1986}
\bibinfo{author}{\bibfnamefont{T.}~\bibnamefont{Kennedy}} \bibnamefont{and}
  \bibinfo{author}{\bibfnamefont{E.}~\bibnamefont{Lieb}},
  \bibinfo{journal}{Physica} \textbf{\bibinfo{volume}{138A}},
  \bibinfo{pages}{320} (\bibinfo{year}{1986}).

\bibitem[{\citenamefont{Brandt and Mielsch}(1989)}]{brandt_mielsch_1989}
\bibinfo{author}{\bibfnamefont{U.}~\bibnamefont{Brandt}} \bibnamefont{and}
  \bibinfo{author}{\bibfnamefont{C.}~\bibnamefont{Mielsch}},
  \bibinfo{journal}{Z. Phys. B} \textbf{\bibinfo{volume}{75}},
  \bibinfo{pages}{365} (\bibinfo{year}{1989}).

\bibitem[{\citenamefont{Brandt and Mielsch}(1990)}]{brandt_mielsch_1990}
\bibinfo{author}{\bibfnamefont{U.}~\bibnamefont{Brandt}} \bibnamefont{and}
  \bibinfo{author}{\bibfnamefont{C.}~\bibnamefont{Mielsch}},
  \bibinfo{journal}{Z. Phys. B} \textbf{\bibinfo{volume}{79}},
  \bibinfo{pages}{295} (\bibinfo{year}{1990}).

\bibitem[{\citenamefont{Brandt and Mielsch}(1991)}]{brandt_mielsch_1991}
\bibinfo{author}{\bibfnamefont{U.}~\bibnamefont{Brandt}} \bibnamefont{and}
  \bibinfo{author}{\bibfnamefont{C.}~\bibnamefont{Mielsch}},
  \bibinfo{journal}{Z. Phys. B} \textbf{\bibinfo{volume}{82}},
  \bibinfo{pages}{37} (\bibinfo{year}{1991}).

\bibitem[{\citenamefont{Gruber et~al.}(2001)\citenamefont{Gruber, Macris,
  Royer, and Freericks}}]{gruber_macris_royer_2001}
\bibinfo{author}{\bibfnamefont{C.}~\bibnamefont{Gruber}},
  \bibinfo{author}{\bibfnamefont{N.}~\bibnamefont{Macris}},
  \bibinfo{author}{\bibfnamefont{P.}~\bibnamefont{Royer}}, \bibnamefont{and}
  \bibinfo{author}{\bibfnamefont{J.~K.} \bibnamefont{Freericks}},
  \bibinfo{journal}{Phys. Rev. B} \textbf{\bibinfo{volume}{63}},
  \bibinfo{pages}{165111} (\bibinfo{year}{2001}).

\bibitem[{\citenamefont{Jarrell}(1992)}]{jarrell_1992}
\bibinfo{author}{\bibfnamefont{M.}~\bibnamefont{Jarrell}},
  \bibinfo{journal}{Phys. Rev. Lett.} \textbf{\bibinfo{volume}{69}},
  \bibinfo{pages}{168} (\bibinfo{year}{1992}).

\bibitem[{\citenamefont{{M\"uller-Hartmann}}(1989)}]{mueller-hartmann_1989b}
\bibinfo{author}{\bibfnamefont{E.}~\bibnamefont{{M\"uller-Hartmann}}},
  \bibinfo{journal}{Z. Phys. B} \textbf{\bibinfo{volume}{76}},
  \bibinfo{pages}{211} (\bibinfo{year}{1989}).

\end{thebibliography}

\end{document}